\documentclass[preprint,12pt]{elsarticle} 
\usepackage{graphicx} 
\usepackage{amssymb} 
\usepackage{amsmath} 

\def\lamb#1#2{$^{#1}_{\Lambda}${#2}}

\journal{Nuclear Physics A} 

\begin{document} 

\begin{frontmatter} 

\title{Charge symmetry breaking in the $A=4$ hypernuclei} 
\author[a,b]{Daniel Gazda} 
\author[c]{Avraham Gal} 
\address[a]{Department of Physics, Chalmers University of Technology, 
SE-412 96 G\"{o}teborg, Sweden} 
\address[b]{Nuclear Physics Institute, 25068 \v{R}e\v{z}, Czech Republic} 
\address[c]{Racah Institute of Physics, The Hebrew University, 91904 
Jerusalem, Israel} 

\begin{abstract} 
Charge symmetry breaking (CSB) in the $\Lambda$-nucleon strong interaction 
generates a charge dependence of $\Lambda$ separation energies in mirror 
hypernuclei, which in the case of the $A=4$ mirror hypernuclei $0^+$ ground 
states is sizable, $\Delta B^{J=0}_{\Lambda}\equiv B^{J=0}_{\Lambda}
(_{\Lambda}^4{\rm He})-B^{J=0}_{\Lambda}(_{\Lambda}^4{\rm H})=230\pm 90$~keV, 
and of opposite sign to that induced by the Coulomb repulsion in 
light hypernuclei. Recent {\it ab initio} calculations of the 
(\lamb{4}{H}, \lamb{4}{He}) mirror hypernuclei $0^+_{\rm g.s.}$ and 
$1^+_{\rm exc}$ levels have demonstrated that a $\Lambda - \Sigma^0$ mixing 
CSB model due to Dalitz and von Hippel (1964) is capable of reproducing this 
large value of $\Delta B^{J=0}_{\Lambda}$. These calculations are discussed 
here with emphasis placed on the leading-order chiral EFT hyperon-nucleon 
Bonn-J\"{u}lich strong-interaction potential model used and the no-core 
shell-model calculational scheme applied. The role of one-pion exchange in 
producing sizable CSB level splittings in the $A=4$ mirror hypernuclei is 
discussed.

\end{abstract}

\begin{keyword} 

hypernuclei, hyperon-nucleon interactions, charge symmetry breaking 


\end{keyword} 

\end{frontmatter}

\section{Introduction} 
\label{sec:intro} 

\begin{figure}[hbt] 
\begin{center} 
\includegraphics[width=0.7\textwidth]{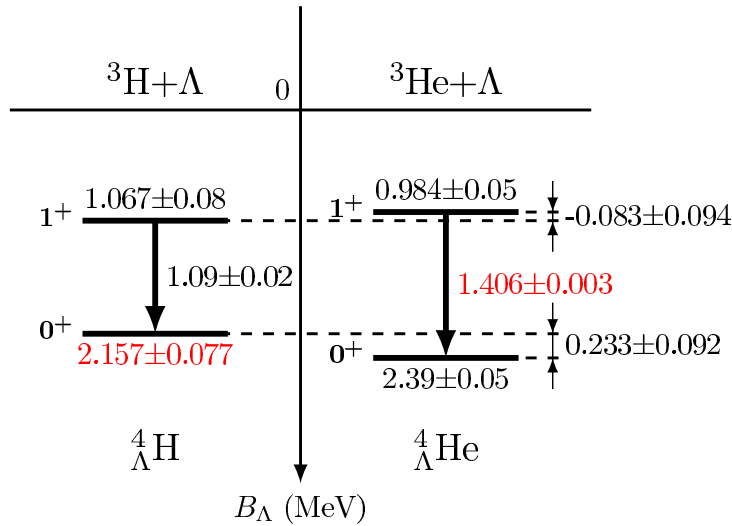}  
\caption{(${_{\Lambda}^4{\rm H}},{_{\Lambda}^4{\rm He}}$) level diagram 
(in MeV). The $0^+_{\rm g.s.}$ $\Lambda$ separation energies $B_{\Lambda}$, 
loosely termed $\Lambda$ binding energies, taken from a recent measurement 
at MAMI \cite{MAMI16} for \lamb{4}{H} and from emulsion work \cite{davis05} 
for \lamb{4}{He}, are marked under the $0^+_{\rm g.s.}$ energy levels. The 
$1^{+}_{\rm exc}$ separation energies follow from $\gamma$-ray measurements 
of the excitation energies $E_{\gamma}$ \cite{E13} denoted by arrows, and are 
marked above the $1^{+}_{\rm exc}$ energy levels. CSB splittings are shown to 
the right of the \lamb{4}{He} levels. Results from recent measurements are 
highlighted in red in the online version. Figure adapted from \cite{MAMI16}.} 
\label{fig:A=4} 
\end{center} 
\end{figure}

Charge symmetry breaking (CSB) in the $\Lambda N$ interaction, which amounts 
to the difference between the $\Lambda n$ and the $\Lambda p$ interactions, 
cannot be studied in free space for lack of direct or indirect $\Lambda n$ 
scattering data and also because none of the two possible $I=\frac{1}{2}$ 
$\Lambda n$ and $\Lambda p$ systems is bound. Furthermore, it cannot be 
inferred from the only three-body $\Lambda$ hypernucleus known to date, 
the $I=0$ \lamb{3}{H}, in which CSB effects are highly suppressed. However, 
the two four-body $I=\frac{1}{2}$ $\Lambda$ hypernuclei, \lamb{4}{H} with 
$I_z=-\frac{1}{2}$ and \lamb{4}{He} with $I_z=+\frac{1}{2}$, each one with 
two particle-stable levels $0^+_{\rm g.s.}$ and $1^+_{\rm exc}$, suggest 
substantial CSB splitting of the $A=4$ hypernuclear ground state 
(see Fig.~\ref{fig:A=4}):  
\begin{equation} 
\Delta B^{J=0}_{\Lambda} \equiv B^{J=0}_{\Lambda}(_{\Lambda}^4{\rm He}) - 
B^{J=0}_{\Lambda}(_{\Lambda}^4{\rm H}) = 233\pm 92~{\rm keV}. 
\label{eq:J=0} 
\end{equation} 

\begin{figure}[hbt] 
\begin{center} 
\includegraphics[width=0.7\textwidth]{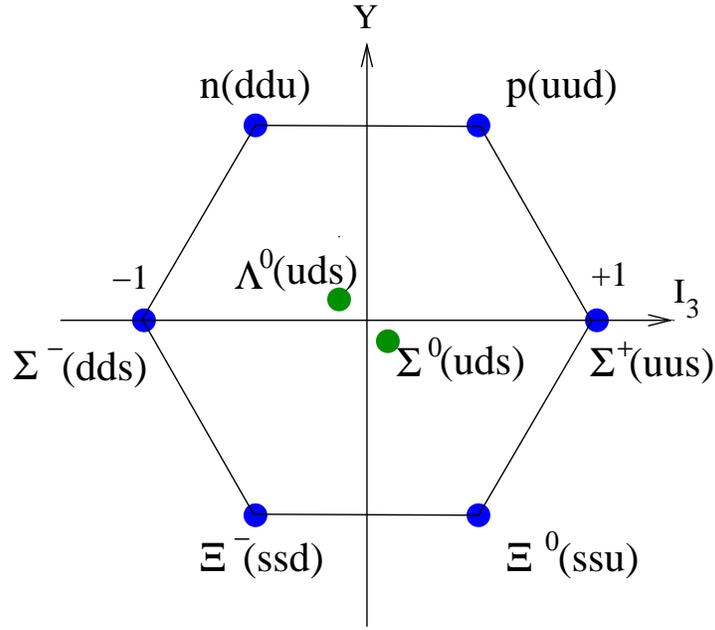} 
\caption{SU(3)$_{\rm f}$ octet baryons with their underlying leading quark 
structure. Note the $I=0$ $\Lambda^0(uds)$ and $I=1$ $\Sigma^0(uds)$ hyperons, 
sharing the $I_z=0$ central point, which are admixed by CSB in the physical 
$\Lambda$ and $\Sigma^0$ hyperons.} 
\label{fig:octet} 
\end{center} 
\end{figure}

Until recently, this relatively large observed CSB splitting could not be 
reproduced in \textit{ab-initio} four-body calculations with the widely used 
hyperon-nucleon ($YN$) Nijmegen soft-core meson exchange models NSC97$_{
\rm e,f}$~\cite{NSC97}; see Refs.~\cite{haidenbauer07,nogga13,nogga14}. 
A maximal value of $\Delta B^{J=0}_{\Lambda}\approx 100$~keV was reached 
in model NSC97$_{\rm f}$ \cite{haidenbauer07}. The CSB model used in these 
past calculations is the $\Lambda-\Sigma^0$ mixing model of Dalitz and von 
Hippel \cite{DvH64}. In this model, the pure-isospin $I=0$ $\Lambda^0(uds)$ 
and $I=1$ $\Sigma^0(uds)$ octet hyperons which share the $I_z=0$ central point 
of the SU(3)$_{\rm f}$ octet, as shown in Fig.~\ref{fig:octet}, are admixed 
by CSB in forming the physical $\Lambda$ and $\Sigma^0$ hyperons. The model 
relates then the mass-mixing matrix element $\langle\Sigma^0|\delta M|\Lambda
\rangle$ to electromagnetic mass differences of SU(3)$_{\rm f}$ octet baryons:  
\begin{equation} 
\langle\Sigma^0|\delta M|\Lambda\rangle=\frac{1}{\sqrt 3}\,
[(M_{\Sigma^0}-M_{\Sigma^+})-(M_n-M_p)]=1.143\pm 0.040~{\rm MeV}.    
\label{eq:deltaM} 
\end{equation} 
Lattice QCD calculations yield so far only half of this value for the 
mass-mixing matrix element \cite{lqcd15a,galprd15}. The reason apparently 
is the omission of QED from these calculations \cite{lqcd15b}. 

\begin{figure}[t!] 
\begin{center} 
\includegraphics[width=0.7\textwidth]{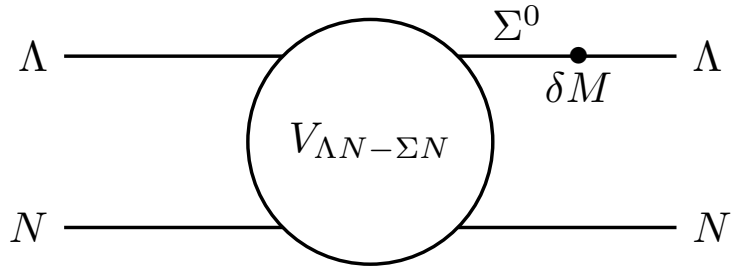} 
\caption{CSB $\Lambda N$ interaction diagram describing a SI $V_{\Lambda N - 
\Sigma N}$ interaction followed by a CSB $\Lambda-\Sigma^0$ mass-mixing 
vertex.} 
\label{fig:diag_csbls_h} 
\end{center} 
\end{figure} 

The mass-mixing matrix element (\ref{eq:deltaM}) serves as insertion in 
$\Lambda N$ CSB diagrams generated by the $\Lambda N\leftrightarrow\Sigma N$ 
strong-interaction (SI) coupling potential $V_{\Lambda N - \Sigma N}$, 
as shown in Fig.~\ref{fig:diag_csbls_h}, leading to a concrete expression 
of $V_{\rm CSB}$ $\Lambda N$ matrix elements in terms of $V_{\rm SI}$ 
$\Lambda N\leftrightarrow\Sigma N$ matrix elements \cite{gal15}: 
\begin{equation} 
\langle N\Lambda|V_{\rm CSB}|N\Lambda\rangle = -0.0297\,\tau_{Nz}\,
\frac{1}{\sqrt{3}}\,\langle N\Sigma|V_{\rm SI}|N\Lambda\rangle , 
\label{eq:OME} 
\end{equation} 
where the z component of the isospin Pauli matrix ${\vec\tau}_N$ assumes the 
values $\tau_{Nz}=\pm 1$ for protons and neutrons, respectively, the isospin 
Clebsch-Gordan coefficient $1/\sqrt{3}$ accounts for the $N\Sigma^0$ amplitude 
in the $I_{NY}=\frac{1}{2}$ $N\Sigma$ state, and the space-spin structure of 
this $N\Sigma$ state is taken identical with that of the $N\Lambda$ state 
embracing $V_{\rm CSB}$. The CSB scale coefficient 0.0297 in (\ref{eq:OME}) 
follows from the $\Lambda-\Sigma^0$ mass-mixing matrix element 
$\langle\Sigma^0|\delta M|\Lambda\rangle$ given above, 
\begin{equation} 
\frac{2\,\langle\Sigma^0|\delta M|\Lambda\rangle}{M_{\Sigma^0}-M_{\Lambda}}=
0.0297\pm 0.0010, 
\label{eq:scale} 
\end{equation} 
where the factor 2 accounts for the two possibilities of inserting $\delta M$ 
in Fig.~\ref{fig:diag_csbls_h}, to the left of the SI circle or to its right 
(as drawn). 

\begin{figure}[hbt]
\begin{center}
\includegraphics[scale=0.5]{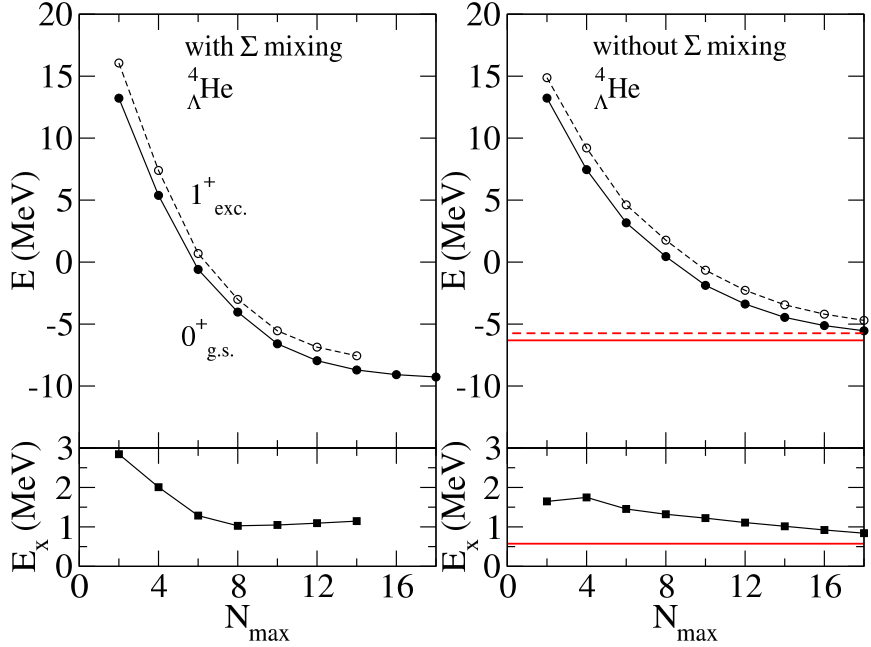}
\caption{Energy eigenvalues $E$ and excitation energies $E_{\rm x}$ in NCSM
calculations of $_{\Lambda}^4{\rm He}(0^+_{\rm g.s.},1^+_{\rm exc})$ states
\cite{gazda14,wirth14} as a function of $N_{\rm max}$, using LO $\chi$EFT
$YN$ interactions with cutoff 600~MeV \cite{polinder06}, including (left) or
excluding (right) $\Lambda\Sigma$ coupling.}
\label{fig:LO}
\end{center}
\end{figure}

Since the charge symmetric SI $\Lambda N\leftrightarrow\Sigma N$ coupling, 
according to Eq.~(\ref{eq:OME}), is the chief provider of the CSB $\Lambda N$ 
matrix element, it is natural to ask how strong the $\Lambda N\leftrightarrow
\Sigma N$ coupling is in realistic microscopic $YN$ interaction models. In 
Fig.~\ref{fig:LO} we show results of no-core shell-model (NCSM) calculations 
of \lamb{4}{He} levels \cite{gazda14,wirth14}, using the Bonn-J\"{u}lich 
leading-order (LO) chiral effective field theory ($\chi$EFT) $YN$ SI potential 
model \cite{polinder06}, in which $\Lambda N\leftrightarrow\Sigma N$ coupling 
is seen to contribute between 3 to 4~MeV to the total binding of \lamb{4}{He} 
and almost 40\% of the $0^+_{\rm g.s.}\to 1^+_{\rm exc}$ excitation energy 
$E_{\rm x}$. A similar effect on $E_{\rm x}$ also occurs in the Nijmegen NSC97 
models \cite{NSC97}. Recall that in a meson exchange model, one-pion exchange 
(OPE), forbidden by isospin in the SI $\Lambda N$ diagonal potential, 
contributes as strongly as possible to the $\Lambda N\leftrightarrow\Sigma N$ 
coupling potential. With SI $\Lambda N\leftrightarrow\Sigma N$ potential 
energy contributions of order 10~MeV \cite{nogga01}, and with a CSB scale of 
order 3\%, Eq.~(\ref{eq:OME}) could yield CSB contributions of order 300~keV. 
As shown below, the Bonn-J\"{u}lich LO $\chi$EFT $YN$ interaction potentials 
\cite{polinder06} are able to produce this order of magnitude by applying 
Eq.~(\ref{eq:OME}) to each one of the $\Lambda N\leftrightarrow\Sigma N$ 
$V_{\rm SI}$ components in this LO version. Disregarded in this procedure 
are CSB contributions arising from meson mixings, such as $\pi^0 - \eta$ 
and $\rho^0 - \omega$. These were found negligible, 
\begin{equation} 
\Delta B^{J=0}_{\Lambda}(\pi^0\eta+\rho^0\omega)\sim -20~{\rm keV},\,\,\, 
\Delta B^{J=1}_{\Lambda}(\pi^0\eta+\rho^0\omega)\sim -10~{\rm keV}, 
\label{eq:coon} 
\end{equation} 
in four-body $\Lambda NNN$ calculations by Coon et al. \cite{coon99} 
and are disregarded here.{\footnote{In particular, correcting an 
oversight in Ref.~\cite{DvH64}, the $\pi^0 - \eta$ mixing contribution 
to $\Delta B^{J=0}_{\Lambda}$ is opposite in sign to the positive $\pi^0$ 
exchange contribution from $\Lambda - \Sigma^0$ mixing \cite{coon79}.}} 

The present work extends our Letter report \cite{gg16} on CSB level-splitting 
calculations in the $A=4$ mirror hypernuclei, adding calculational details, 
and furthermore comparing the CSB splittings derived from these {\it ab 
initio} calculations with those derived by a straightforward evaluation of OPE 
CSB contributions. The paper is organized as follows: in Sect.~\ref{sec:meth} 
we review briefly the Bonn-J\"{u}lich LO $\chi$EFT approach followed in our 
NCSM four-body calculations, as well as providing details of the application 
of this NCSM technique. Results of these calculations, updating and extending 
those of Ref.~\cite{gg16}, are presented in Sect.~\ref{sec:res}, with further 
discussion centered on the role of OPE in Sect.~\ref{sec:disc}. The paper 
ends with a brief summary and outlook in Sect.~\ref{sec:sum}.

\section{Methodology} 
\label{sec:meth}

\subsection{NCSM hypernuclear calculations} 
\label{subsec:NCSM} 

The version of the NCSM approach which is particularly suitable for dealing 
with few-body systems employs translationally invariant harmonic-oscillator 
(HO) bases formulated in relative Jacobi coordinates \cite{navratil00} 
in which two-body and three-body interaction matrix elements are evaluated. 
Antisymmetrization is imposed with respect to nucleons, and the resulting 
Hamiltonian is diagonalized in a finite four-body HO basis, admitting all HO 
excitation energies $N\hbar\omega$, $N\leq N_{\rm max}$, up to $N_{\rm max}$ 
HO quanta. 

This NCSM nuclear technique was extended recently to light hypernuclei 
\cite{gazda14,wirth14} and is applied here in a particle basis, with full 
account of the different masses within baryon iso-multiplets, to the 
\lamb{4}{H} and \lamb{4}{He} mirror hypernuclei, using momentum-space 
chiral model interactions specified in Sect.~\ref{subsec:LO}. Some technical 
details of the present application of the NCSM methodology to the $A$=4 
mirror hypernuclei are relegated to the unpublished Appendix A. While it was 
possible to obtain fully converged binding energies, with keV precision, 
for the $A$=3 core nuclei $^3$H and $^3$He, it was not computationally 
feasible to perform calculations with sufficiently large $N_\text{max}$ to 
demonstrate convergence for \lamb{4}{H} and \lamb{4}{He}. In these cases 
extrapolation to an infinite model space, $N_\text{max}\rightarrow\infty$, 
had to be employed.{\footnote{The issue of extrapolation in NCSM is 
unsettled, with somewhat inconclusive discussions of error estimates; 
see e.g.\ Refs.~\cite{maris09,wendt15,liebig15,coon16} and work cited 
therein for more elaborate methods than the ones employed here.}} 
Extrapolated energy values $E(\omega)$ are obtained in the present 
work by fitting an exponential function, 
\begin{equation}
\label{eq:expfit}
E(N_\text{max},\omega)=E(\omega)+A\, \text{e}^{-B\, N_{\max}}, 
\end{equation}
with parameters $A$ and $B$, to $E(N_{\rm max},\omega \text{~fixed})$ 
sequences in the vicinity of the variational minima with respect to the HO 
basis frequency $\omega$. The reliability of such extrapolations is then 
reflected in the independence of $E(\omega)$ of the frequency $\omega$. 
In our fitting procedure, only the last three $N_\text{max}$ values which 
are the most reliable ones, were used. 

\begin{figure}[hbt]
\begin{center}
\includegraphics[width=0.7\textwidth]{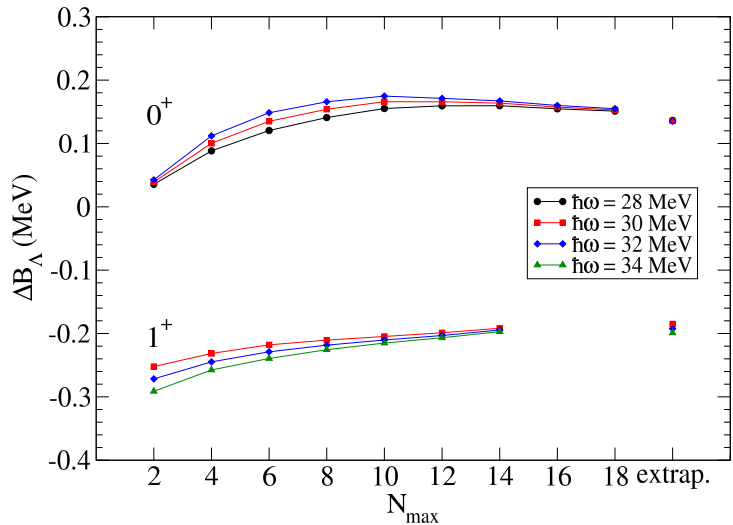}
\caption{Dependence of the separation-energy differences $\Delta B_{\Lambda}$
between $_{\Lambda}^4{\rm He}$ and $_{\Lambda}^4{\rm H}$, for $0^+_{\rm g.s.}$
(upper curves) and for $1^+_{\rm exc}$ (lower curves) on the model-space size 
parameter $N_\text{max}$ for HO values of $\hbar\omega$ around the variational 
minima, together with their extrapolated values, in {\it ab initio} NCSM 
calculations using LO chiral EFT coupled-channel $YN$ potentials $V_{\rm SI}$, 
with cutoff momentum $\Lambda$=600~MeV \cite{polinder06}, plus $V_{\rm CSB}$ 
derived from $V_{\rm SI}$ using Eq.~(\ref{eq:OME}).}
\label{fig:csbnmaxom}
\end{center}
\end{figure}

It is worth noting that the present work focuses on \emph{differences} 
$\Delta B_{\Lambda}$ of $\Lambda$-hyperon separation energies in \lamb{4}{H} 
and \lamb{4}{He}, obtained as 
\begin{equation}
\label{eq:deltab}
\Delta B_{\Lambda}=[E({}^3\text{He})-E({}^4_\Lambda\text{He})]-
[E({}^3\text{H})-E({}^4_\Lambda\text{H})], 
\end{equation}
where converged energy values of $^3$H and $^3$He are used together with 
extrapolated energy values for \lamb{4}{H} and \lamb{4}{He}. In general, 
the differences $\Delta B_\Lambda$ are much more stable as function of 
$N_\text{max}$ than the absolute energies are. This is demonstrated in 
Fig.~\ref{fig:csbnmaxom} where the dependence of the separation-energy 
differences $\Delta B_{\Lambda}$, for the $0^+_{\rm g.s.}$ (upper curves) 
and the $1^+_{\rm exc}$ (lower curves) states, on the size of the model 
space is shown for HO frequencies $\omega$ around the variational minima of 
absolute energies at $\hbar\omega=30(32)$~MeV for $J$=0(1), together with 
their extrapolated values. The values of $\Delta B_{\Lambda}$ exhibit fairly 
weak $N_\text{max}$ and $\omega$ dependence compared to the behavior of 
the absolute energies, and to a lesser extent the behavior of the $\Lambda$ 
separation energies, and the employed extrapolation scheme is found 
sufficiently robust for our purposes. With regard to the use of $N_\text{max}
\to\infty$ extrapolated values based on the last three $N_\text{max}$ values, 
it was found that including the last four $N_\text{max}$ values in the fit 
resulted in $\Delta B_\Lambda$ values that differed by $\lesssim 10$~keV.

\subsection{LO $\chi$EFT $YN$ interaction input} 
\label{subsec:LO} 

$\chi$EFT interactions are used throughout this work, with N3LO $NN$ and 
N2LO $NNN$ interactions, \cite{entem03,navratil07} respectively, both 
with momentum cutoff $\Lambda=500$~MeV. For the SI $YN$ coupled-channel 
potentials $V_{\rm SI}$ we use the Bonn-J\"{u}lich SU(3)-based LO $\chi$EFT 
approach \cite{polinder06} plus $V_{\rm CSB}$ evaluated from $V_{\rm SI}$ by 
using Eq.~(\ref{eq:OME}). In principle, the power counting underlying the EFT 
scheme allows to include two $\Lambda N$ CSB contact terms, as done in N3LO 
$NN$ versions to account quantitatively for the charge dependence of the 
low-energy $NN$ scattering parameters \cite{entem03,epelbaum05}. Given, 
however, that low-energy $\Lambda p$ cross sections are poorly known and 
$\Lambda n$ scattering data are unavailable, the corresponding low-energy 
constants cannot be determined, unless they are fitted to the two CSB 
splittings $\Delta B^{J=0,1}_{\Lambda}$ of the $A=4$ hypernuclear mirror 
levels, in which case the $A=4$ CSB calculation reduces to tautology. This 
unfortunate occurrence cannot be remedied by going from LO to NLO $\chi$EFT 
$YN$ potentials. For this reason, and anticipating that hypernuclear CSB is 
driven by the relatively long-range OPE, we disregard CSB contact terms. 

The $\chi$EFT potentials $V_{\rm SI}$ are regularized in momentum space, 
using the standard choice \cite{haidenbauer07} 
\begin{equation} 
\langle p'|V_{\rm SI}|p \rangle \rightarrow \langle p'|V_{\rm SI}|p \rangle 
\times \exp (-\frac{p'^4+p^4}{\Lambda^4}), 
\label{eq:cutoff} 
\end{equation} 
in order to remove high-energy components of the hadronic fields 
involved.{\footnote{Unfortunately such momentum-space non-local regulators 
affect also the long-range part of the potentials, as noted recently by 
Epelbaum et al. \cite{epelbaum15} who advocated using coordinate-space local 
regulators.}} 
In LO, $V_{\rm SI}$ consists of regularized pseudoscalar (PS) $\pi$, $K$ and 
$\eta$ meson exchanges with coupling constants constrained by SU(3)$_{\rm f}$, 
plus five central interaction low-energy constants (also called contact terms) 
simulating the short range behavior of the $YN$ coupled channel interactions, 
all of which are regularized according to (\ref{eq:cutoff}) with a cutoff 
momentum $\Lambda\geq m_{\rm PS}$, varied from 550 to 700 MeV. Two of the five 
contact terms connect $\Lambda N$ to $\Sigma N$ in spin-singlet and triplet 
$s$-wave channels, and are of special importance for the calculation of CSB 
splittings. The dominant meson exchange interaction is OPE which couples the 
$\Lambda N$ channel exclusively to the $I=\frac{1}{2}$ $\Sigma N$ channel. 
$K$-meson exchange also couples these two $YN$ channels. This LO $V_{\rm SI}$ 
($V_{\rm SI}^{\rm LO}$) reproduces reasonably well, with $\chi^2/(\rm{d.o.f.})
\approx 1$, the scarce $YN$ low-energy scattering data. 
It also reproduces the binding energy of \lamb{3}{H}, with a calculated value 
$B_{\Lambda}$(\lamb{3}{H})=110$\pm$10~keV for cutoff 600 MeV \cite{wirth14}, 
consistent with experiment (130$\pm$50~keV \cite{davis05}) and with Faddeev 
calculations reported by Haidenbauer et al. \cite{haidenbauer07}. Isospin 
conserving matrix elements of $V_{\rm SI}^{\rm LO}$, given in momentum 
space, are evaluated here in a particle basis with full account of mass 
differences within baryon iso-multiplets, while isospin breaking $I_{NN}$ 
$0\leftrightarrow 1$ and $I_{YN}$ $\frac{1}{2}\leftrightarrow \frac{3}{2}$ 
transitions are suppressed. The Coulomb interaction between charged baryons 
($pp$, $\Sigma^{\pm}p$) is included. 

Calculations consisting of fully converged $^3$H and $^3$He binding 
energies, and (\lamb{4}{H}, \lamb{4}{He}) $0^+_{\rm g.s.}$ and 
$1^{+}_{\rm exc}$ binding energies extrapolated to infinite model spaces 
from $N_{\rm max}=18(14)$ for $J=0(1)$ are reported in the next section. 
The calculated binding energies of the core nuclei, 8.482 MeV for $^3$H 
and 7.720 MeV for $^3$He, reproduce very well the known binding energies.
The $NNN$ interaction, was excluded from most of the hypernuclear 
calculations after verifying that, in spite of adding almost 80 keV to 
the $\Lambda$ separation energies $B^{J=0}_{\Lambda}$ and somewhat less to 
$B^{J=1}_{\Lambda}$, its inclusion makes a difference of only a few keV for 
the CSB splittings $\Delta B^{J}_{\Lambda}$ in both the $0^+_{\rm g.s.}$ 
and $1^+_{\rm exc}$ states.

\section{Results} 
\label{sec:res} 

This section is divided to two parts, one in which the explicit CSB potential 
$V_{\rm CSB}$ of Eq.~(\ref{eq:OME}) is excluded, in order to allow comparison 
with past calculations, and one in which $V_{\rm CSB}$ is generated from the 
LO $\chi$EFT $\Lambda N \leftrightarrow \Sigma N$ strong interactions used 
here. 

\subsection{Without explicit CSB} 
\label{subsec:noCSB} 

We start by comparing in Table~\ref{tab:B_L(L4H0+noCSB)} ($0^+_{\rm g.s.}$) 
and Table~\ref{tab:B_L(L4H1+noCSB)} ($1^+_{\rm exc}$) our NCSM calculations to 
Nogga's Yakubovsky-equations calculations for \lamb{4}{H}, both using the same 
LO $\chi$EFT $YN$ interactions with no explicit CSB potential $V_{\rm CSB}$, 
and also to Nogga's recent calculations using NLO \cite{nogga13}. With 
uncertainties in calculated $B_{\Lambda}$ values arising from different 
$NN$ input in different LO calculations, and also from the suppressed $NNN$ 
interaction, all of which are conservatively estimated to be of the order 
of $\sim$0.1~MeV, we cite LO results up to the first decimal point. 

\begin{table}[hbt]
\caption{$B^{J=0}_{\Lambda}$(\lamb{4}{H}) values calculated in $\chi$EFT 
approaches, without explicit $V_{\rm CSB}$, for various cutoff momenta 
$\Lambda$ (in MeV). ${\overline B}^{J=0}_{\Lambda}$(\lamb{4}{H}) stands 
for the mean$\pm$spread of these values.} 
\begin{tabular}{lccccc} 
\hline 
$YN$ chiral model & $\Lambda$=550 & $\Lambda$=600 & $\Lambda$=650 & 
$\Lambda$=700 & ${\overline B}^{J=0}_{\Lambda}$(\lamb{4}{H}) \\ 
\hline 
LO (present) & 2.6 & 2.4 & 2.2 & 2.3 & 2.4$\pm$0.2 \\ 
LO (Nogga \cite{nogga13}) & 2.6 & 2.5 & 2.4 & 2.4 & 2.5$\pm$0.1 \\ 
NLO (Nogga \cite{nogga13}) &1.52&1.47&1.52&1.61&1.53$^{+0.08}_{-0.06}$ \\
\hline 
\end{tabular} 
\label{tab:B_L(L4H0+noCSB)} 
\end{table} 

With estimated NCSM $N_{\rm max}\to\infty$ extrapolation uncertainties 
$\pm$0.1~MeV for $0^+_{\rm g.s.}$, Table~\ref{tab:B_L(L4H0+noCSB)} 
demonstrates a very good agreement between the two LO calculations for 
$0^+_{\rm g.s.}$ over the full range of momentum cutoff $\Lambda$ values. 
Both LO calculations exhibit a moderate cutoff dependence, quantified here by 
giving the spread of the cutoff-dependent $B^{J=0}_{\Lambda}$(\lamb{4}{H}) 
values around their mean value. The mean value in ${\overline B}^{J=0}_{
\Lambda}$(\lamb{4}{H}) is close to that expected for \lamb{4}{H} once 
a negative CSB contribution of the order of $\sim$100~keV is added. 
It is worth noting that while the cutoff dependence at NLO is remarkably weak, 
the calculated $B^{J=0}_{\Lambda}$(\lamb{4}{H}) values fall substantially 
below the $0^+_{\rm g.s.}$ experimental separation energy. This could 
signal a need to introduce $YNN$ three-body terms, as suggested recently by 
Petschauer et al. \cite{petschauer16}. However, as argued by us in the Letter 
version of this work \cite{gg16}, these terms are unlikely to give rise to 
additional CSB contributions. 
 
\begin{table}[hbt]
\caption{$B^{J=1}_{\Lambda}$(\lamb{4}{H}) values calculated in $\chi$EFT 
approaches, without explicit $V_{\rm CSB}$, for various cutoff momenta 
$\Lambda$ (in MeV). ${\overline B}^{J=1}_{\Lambda}$(\lamb{4}{H}) stands 
for the mean$\pm$spread of these values.}  
\begin{tabular}{lccccc} 
\hline 
$YN$ chiral model & $\Lambda$=550 & $\Lambda$=600 & $\Lambda$=650 & 
$\Lambda$=700 & ${\overline B}^{J=1}_{\Lambda}$(\lamb{4}{H}) \\  
\hline 
LO (present) & 1.7 & 1.3 & 0.9 & 0.5 & 1.1$\pm$0.6 \\ 
LO (Nogga \cite{nogga13}) & 1.9 & 1.5 & 1.2 & 1.0 & 1.4$^{+0.5}_{-0.4}$ \\ 
NLO (Nogga \cite{nogga13}) & 0.85&0.73&0.83& 0.90 & 0.83$^{+0.07}_{-0.10}$ \\ 
\hline 
\end{tabular} 
\label{tab:B_L(L4H1+noCSB)} 
\end{table} 

Table~\ref{tab:B_L(L4H1+noCSB)} exhibits a much stronger cutoff dependence 
of $B^{J=1}_{\Lambda}$(\lamb{4}{H}) in both LO calculations. Our NCSM 
$N_{\rm max}\to\infty$ extrapolation uncertainties, estimated as $\pm$0.5~MeV 
for the $1^+_{\rm exc}$ state, are considerably larger than for the 
$0^+_{\rm g.s.}$, reflecting perhaps the weaker binding of the excited 
state as also noted in Nogga's work \cite{nogga01}. Given these uncertainties, 
the table demonstrates, again, a reasonable agreement between the two LO 
calculations. In contrast, the NLO $B^{J=1}_{\Lambda}$ values show a very 
weak cutoff dependence, as weak almost as for the $0^+_{\rm g.s.}$ in NLO, 
but all of these $B^{J=1}_{\Lambda}$ values fall considerably below that 
anticipated from the $1^+_{\rm exc}$ experimental separation energy. 
This might suggest, again, a need to introduce $YNN$ three-body terms. 

\begin{table}[hbt]
\caption{$E_{\rm x}(0^+_{\rm g.s.}\to 1^+_{\rm exc})$ in \lamb{4}{H} 
calculated in $\chi$EFT approaches without explicit $V_{\rm CSB}$ for 
various cutoff momenta $\Lambda$ (in MeV).} 
\begin{tabular}{lccccc} 
\hline 
$YN$ chiral model & $\Lambda$=550 & $\Lambda$=600 & $\Lambda$=650 & 
$\Lambda$=700 & ${\overline E}_{\rm x}$(\lamb{4}{H}) \\ 
\hline 
LO (present) & 0.9 & 1.1 & 1.3 & 1.8 & 1.3$^{+0.5}_{-0.4}$ \\ 
LO (Nogga \cite{nogga13}) & 0.8 & 1.0 & 1.1 & 1.3 & 1.05$\pm$0.25 \\ 
NLO (Nogga \cite{nogga13}) & 0.67 & 0.75 & 0.69 & 0.71 & 0.71$\pm$0.04 \\ 
\hline 
\end{tabular} 
\label{tab:Ex(L4HnoCSB)} 
\end{table} 

The underbinding noted above for the NLO results is manifest also upon 
inspecting the calculated excitation energies $E_{\rm x}(0^+_{\rm g.s.}\to 
1^+_{\rm exc})$ listed in Table~\ref{tab:Ex(L4HnoCSB)}. Whereas both LO 
calculations reproduce the value of $E_{\rm x}$ expected from experiment, 
albeit by virtue of the large spread of their $\Lambda$ dependent $E_{\rm x}$ 
values, the nearly $\Lambda$-independent $E_{\rm x}$ values in NLO are short 
by roughly 0.4$\pm$0.1~MeV of reproducing the value expected from experiment. 

\begin{table}[hbt]
\caption{Cutoff dependence of $A=4$ hypernuclear mirror-level splittings
$\Delta B^J_{\Lambda}(A=4)$ (in keV) from {\it ab initio} NCSM calculations,
using LO $YN$ \cite{polinder06} and N3LO $NN$ \cite{entem03} $\chi$EFT strong 
interactions plus Coulomb interactions, without any explicit $V_{\rm CSB}$.
The HO $\hbar\omega$ values used are 32~MeV for cutoffs $\Lambda=550,600$~MeV
and 34~MeV for $\Lambda=650,700$~MeV.}
\begin{tabular}{lcccc}
\hline
$\Delta B^J_{\Lambda}(A=4)$ & $\Lambda$=550 MeV & $\Lambda$=600 MeV &
$\Lambda$=650 MeV & $\Lambda$=700 MeV \\
\hline
$J=0$ (keV) & $-$37 & $-$9 & $+$6 & $+$19 \\
$J=1$ (keV) & $-$52 & $-$46 & $-$31 & $-$25 \\
\hline
\end{tabular}
\label{tab:CS}
\end{table}

Although no explicit CSB potential $V_{\rm CSB}$ was used in the calculations 
briefed in this subsection, small residual CSB splittings of hypernuclear 
mirror levels arise, mainly from two sources: (i) the increased repulsive 
Coulomb energy of \lamb{4}{He} with respect to that of its $^3$He nuclear 
core, estimated long ago by Bodmer and Usmani \cite{bodmer85} in a Monte-Carlo 
four-body calculation, 
\begin{equation} 
\Delta B^{J=0}_{\Lambda}({\rm Coul})=-50\pm 20~{\rm keV}, \,\,\,\,\,\, 
\Delta B^{J=1}_{\Lambda}({\rm Coul})=-25\pm 15~{\rm keV}, 
\label{eq:Coulomb} 
\end{equation} 
which for $J=0$ is of opposite sign to the positive $\Delta B^{J=0}_{\Lambda}$ 
observed; and (ii) $\Sigma N$ intermediate-state mass differences in kinetic 
energy terms, estimated by Nogga et al. \cite{nogga02} (see also Table 2 in 
Ref.~\cite{gal15}) for the $0^+_{\rm g.s.}$ as 
\begin{equation} 
\Delta B^{J=0}_{\Lambda}(\Delta M_{\Sigma})\sim \frac{2}{3}\, 
(M_{\Sigma^-}-M_{\Sigma^+})\,P_{\Sigma}\approx 50\pm 10~{\rm keV}, 
\label{eq:DeltaT} 
\end{equation}  
where $P_{\Sigma}$ is the $\Sigma NNN$ admixture probability, of the order of 
1\% in the $0^+_{\rm g.s.}$ and considerably smaller for the $1^+_{\rm exc}$ 
state. There is substantial cancellation between these two contributions as 
seen from Table \ref{tab:CS} where we list differences $\Delta B^J_{\Lambda}
(A=4)$ of separation energies computed at given values of $\omega$ on top 
or near the absolute variational energy minima from $N_{\rm max}=18(14)$ 
output for $J=0(1)$, using LO $\chi$EFT coupled-channel $YN$ potentials 
\cite{polinder06} with no explicit $V_{\rm CSB}$. The uncertainty associated 
with the specific choice of $\omega$ amounts to few keV at most. Since the 
$\Sigma NNN$ admixture probability increases with the cutoff momentum 
$\Lambda$, owing to the small spatial extension of the $\Sigma NNN$ components 
of the four-body wave function, the $\Sigma NNN$ admixture kinetic-energy 
positive contribution gradually (as function of $\Lambda$) takes over the 
long-range Coulomb potential negative contribution in the $0^+_{\rm g.s.}$, 
whereas in the $1^+_{\rm exc}$ state it only reduces the magnitude of the 
latter by about 50\%.

\subsection{With explicit CSB}
\label{subsec:CSB}

\begin{table}[hbt]
\caption{Cutoff dependence of $\Lambda$ separation energies $B^{J}_{\Lambda}$  
in $_{\Lambda}^4{\rm H}$ and $_{\Lambda}^4{\rm He}$ (all in MeV) from 
{\it ab initio} NCSM calculations, using LO $YN$ \cite{polinder06} and N3LO 
$NN$ \cite{entem03} $\chi$EFT strong interactions plus Coulomb interactions, 
and $V_{\rm CSB}$ generated by Eq.~(\ref{eq:OME}) from $V_{\rm SI}^{\rm LO}$. 
Experimental values are from Fig.~\ref{fig:A=4}.}
\begin{tabular}{lccccc} 
\hline 
$B^J_{\Lambda}$(\lamb{4}{Z}) & $\Lambda$=550 & $\Lambda$=600 & $\Lambda$=650 
& $\Lambda$=700 & Experiment \\ 
\hline 
$B^{J=0}_{\Lambda}(_{\Lambda}^4{\rm H})$ & 2.556 & 2.308 & 2.121 & 2.127 & 
2.16$\pm$0.08 \\ 
$B^{J=0}_{\Lambda}(_{\Lambda}^4{\rm He})$ & 2.586 & 2.444 & 2.365 & 2.423 & 
2.39$\pm$0.05 \\ 
$B^{J=1}_{\Lambda}(_{\Lambda}^4{\rm H})$ & 1.744 & 1.359 & 0.920 & 0.738 & 
1.07$\pm$0.08 \\ 
$B^{J=1}_{\Lambda}(_{\Lambda}^4{\rm He})$ & 1.572 & 1.166 & 0.683 & 0.482 & 
0.98$\pm$0.05 \\ 
\hline 
\end{tabular} 
\label{tab:B_L(A=4)} 
\end{table} 

In Table~\ref{tab:B_L(A=4)} we show the cutoff dependence of the calculated 
$\Lambda$ separation energies $B^{J}_{\Lambda}(A=4)$ for the $A$=4 mirror 
hypernuclei, obtained from NCSM calculations with LO $\chi$EFT coupled-channel 
$YN$ potentials \cite{polinder06} and $V_{\rm CSB}$ from Eq.~(\ref{eq:OME}). 
The listed values are derived from $N_{\rm max}\to\infty$ extrapolated 
binding energy values for the $_{\Lambda}^4{\rm He}$ and $_{\Lambda}^4{\rm H}$ 
$J=0,1$ levels at the cutoff-dependent absolute variational minima which are 
$\hbar\omega(J=0)$=30,30,32,34 MeV and $\hbar\omega(J=1)$=32,32,34,36 MeV for 
cutoff values $\Lambda$=550,600,650,700 MeV, respectively. We note that the 
spread of $B^{J}_{\Lambda}(\hbar\omega)$ values near the absolute variational 
minimum for a given cutoff momentum is of the order of 30 keV for $J=0$ and 
considerably larger, about 150~keV, for $J=1$; however, as demonstrated in 
Fig.~\ref{fig:csbnmaxom}, it is considerably smaller, in fact marginal, for 
the CSB splittings $\Delta B^{J}_{\Lambda}$ which are the main topic of the 
present work. 

The $\Lambda$ separation energies listed in Table~\ref{tab:B_L(A=4)} 
show a moderate cutoff dependence for the $0^+_{\rm g.s.}$ mirror levels 
and a stronger dependence for the $1^{+}_{\rm exc.}$ mirror levels, 
with mean values for their charge-symmetric (CS) averages given by 
${\overline B}^{\rm CS}_{\Lambda}(0^+_{\rm g.s.})$=2.37$^{+0.20}_{-0.13}$~MeV 
and ${\overline B}^{\rm CS}_{\Lambda}(1^{+}_{\rm exc.})$=1.08$^{+0.58}_{
-0.47}$~MeV comparing well within their spread with the CS-averaged 
experimental values derived from the last column in the table. Furthermore, 
considering NCSM $N_{\rm max}\to\infty$ extrapolation uncertainties, our 
CS-averaged $B_{\Lambda}$ values are in fair agreement with those reported 
in other four-body calculations using CS LO $YN$ $\chi$EFT interactions 
\cite{haidenbauer07,nogga13,nogga14,gazda14,wirth14}. 

\begin{table}[hbt]
\caption{Cutoff dependence of $A=4$ hypernuclear mirror-level splittings 
$\Delta B^J_{\Lambda}(A=4)$ (in keV) extracted from the $B^J_{\Lambda}$ values 
listed in Table~\ref{tab:B_L(A=4)}. The {\it ab initio} NCSM calculations that 
yield these values use LO $YN$ \cite{polinder06} and N3LO $NN$ \cite{entem03} 
$\chi$EFT interactions plus Coulomb interactions, with $V_{\rm CSB}$ generated 
by Eq.~(\ref{eq:OME}) from the LO SI $YN$ potentials.}
\begin{tabular}{lcccc} 
\hline 
$\Delta B^J_{\Lambda}(A=4)$ & $\Lambda$=550 MeV & $\Lambda$=600 MeV & 
$\Lambda$=650 MeV & $\Lambda$=700 MeV \\ 
\hline 
$J=0$ (keV) & 30 & 136 & 244 & 296  \\  
$J=1$ (keV) & $-$172 & $-$193 & $-$237 & $-$256  \\ 
\hline 
\end{tabular} 
\label{tab:CSB} 
\end{table} 

The $B^J_{\Lambda}$ values listed in Table~\ref{tab:B_L(A=4)} demonstrate 
substantial CSB, particularly for the higher values of the cutoff momentum 
$\Lambda$. The derived CSB level splittings $\Delta B^J_{\Lambda}$ are listed 
in Table~\ref{tab:CSB}. One notes a strong cutoff momentum dependence of 
$\Delta B^{J=0}_{\Lambda}$, varying between 30 to 300~keV upon increasing 
$\Lambda$, together with moderate cutoff dependence of $\Delta B^{J=1}_{
\Lambda}$, varying between $-$170 to $-$260~keV, just the opposite than for 
the separation energies $B^J_{\Lambda}$. Note that $\Delta B^{J=0}_{\Lambda}$ 
comes out invariably positive, whereas $\Delta B^{J=1}_{\Lambda}$ is robustly 
negative. With mean values ${\overline{\Delta B}}^{J=0}_{\Lambda}$=177$^{
+119}_{-147}$~keV and ${\overline{\Delta B}}^{J=1}_{\Lambda}$=$-215^{+43}_{
-41}$~keV, the mean values ${\overline{\Delta B}}^J_{\Lambda}$ satisfy 
\begin{equation} 
{\overline{\Delta B}}^{J=1}_{\Lambda} \approx 
-\,{\overline{\Delta B}}^{J=0}_{\Lambda} < 0. 
\label{eq:minus} 
\end{equation} 

As discussed in our Letter \cite{gg16}, the reason for the opposite signs 
and approximately equal sizes of the $J=0,1$ CSB level splittings is the 
dominance of the $^1S_0$ contact term (CT) in the SI $\Lambda N\leftrightarrow
\Sigma N$ coupling potential of the LO chiral EFT $YN$ Bonn-J\"{u}lich 
approach \cite{polinder06}. The $^3S_1$ CT is completely negligible in this 
LO version, whereas the other contributions to $\Delta B^J_{\Lambda}$, arising 
from PS SU(3)-flavor octet (${\bf 8_{\rm f}}$) meson exchanges, are relatively 
small and of opposite sign to that of the $^1S_0$ CT contribution. 
For $\Lambda=650$~MeV, for example, 
\begin{equation} 
\Delta B^{J=0}_{\Lambda}({\rm CT})=313~{\rm keV}, \,\,\,\,\, 
\Delta B^{J=0}_{\Lambda}({\bf 8_{\rm f}})=-76~{\rm keV}, 
\label{eq:MO0} 
\end{equation}
\begin{equation} 
\Delta B^{J=1}_{\Lambda}({\rm CT})=-354~{\rm keV}, \,\,\,\,\, 
\Delta B^{J=1}_{\Lambda}({\bf 8_{\rm f}})=69~{\rm keV}. 
\label{eq:MO1} 
\end{equation}
Note that the CT and ${\bf 8_{\rm f}}$ splittings listed here do not add up 
precisely to the corresponding total values listed in Table~\ref{tab:CSB} 
owing to the small `background' CSB contributions surviving in the limit 
$V_{\rm  CSB}\to 0$ (see Table~\ref{tab:CS}) which are present in each one 
of the listed $\Delta B^{J}_{\Lambda}$ values. The small PS ${\bf 8_{\rm f}}$ 
meson exchange contributions, including that of the $\pi$ meson, are opposite 
in sign to the Dalitz--von Hippel (DvH) OPE contribution \cite{DvH64}, which 
is known to be the strongest meson exchange among the PS ${\bf 8_{\rm f}}$ 
meson exchanges. We discuss this puzzling situation in the next section. 

\begin{figure}[t!] 
\begin{center} 
\includegraphics[width=0.7\textwidth]{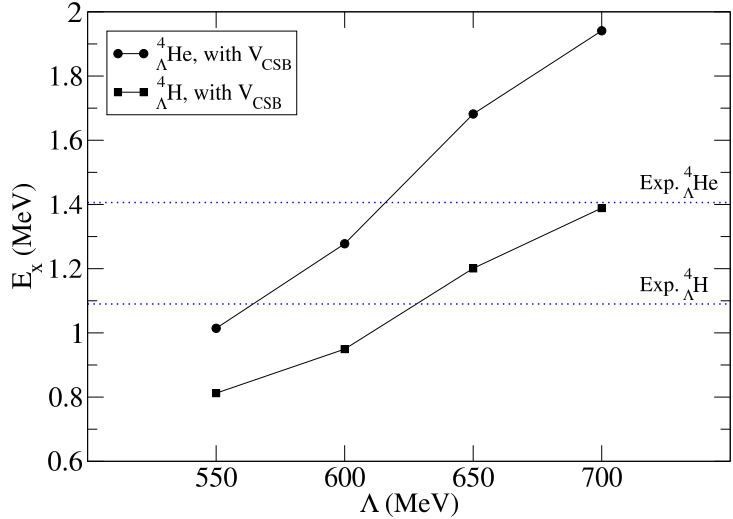} 
\caption{Cutoff momentum dependence of excitation energies 
$E_{\rm x}$(0$^+_{\rm g.s.}$$\to$1$^+_{\rm exc}$) in $_{\Lambda}^4{\rm H}$ 
(squares; lower curve) and $_{\Lambda}^4{\rm He}$ (circles; upper curve) in 
{\it ab initio} NCSM calculations, at $\hbar\omega$ values yielding absolute 
variational minima of the total hypernuclear bound-state energy, for LO chiral 
EFT coupled-channel $YN$ potentials \cite{polinder06} with $V_{\rm CSB}$ 
derived from these SI potentials using Eq.~(\ref{eq:OME}). The dotted 
horizontal lines denote $E_{\rm x}$ values from $\gamma$-ray measurements 
\cite{E13}.} 
\label{fig:Ex} 
\end{center} 
\end{figure}

The next two figures update two similar ones from our Letter \cite{gg16} in 
which the values $\hbar\omega=30(32)$~MeV for $J=0(1)$ were used invariably 
over the full range of values of the cutoff momentum $\Lambda$ spanned in 
these figures. The presently used $\hbar\omega$ values are those for the 
absolute variational energy minima obtained in the NCSM calculations. 
In Fig.~\ref{fig:Ex} we show by solid lines the cutoff momentum dependence 
of the $0^+_{\rm g.s.}\to 1^{+}_{\rm exc}$ excitation energies $E_{\rm x}$ 
formed from the $B_{\Lambda}$ values listed in Table~\ref{tab:B_L(A=4)} 
for both $A$=4 mirror hypernuclei. The dotted horizontal lines mark the 
values of $E_{\rm x}$ deduced from $\gamma$-ray measurements \cite{E13}; 
see Fig.~\ref{fig:A=4}. The crossing of these dotted lines with the respective 
$E_{\rm x}$ solid lines suggests that a choice of cutoff momentum $\Lambda$ 
between 600 and 650 MeV gives the best reproduction of $E_{\rm x}$.  
As noted in several few-body calculations of $s$-shell hypernuclei 
\cite{GL88,akaishi00,hiyama02,akaishi02}, and also demonstrated here in 
Fig.~\ref{fig:LO}, $E_{\rm x}$ is strongly correlated with the $\Lambda N
\leftrightarrow\Sigma N$ coupling potential which in the present context, 
through $\Lambda-\Sigma^0$ mixing, gives rise to CSB splittings of the 
$A=4$ mirror levels. One expects then a similarly strong correlation for 
the CSB splitting of $E_{\rm x}$. Indeed, Fig.~\ref{fig:Ex} shows clearly 
that as $E_{\rm x}$ increases with $\Lambda$, so does the difference 
$\Delta E_{\rm x}\equiv E_{\rm x}$(\lamb{4}{He})$- E_{\rm x}$(\lamb{4}{H}). 

\begin{figure}[t!] 
\begin{center} 
\includegraphics[width=0.7\textwidth]{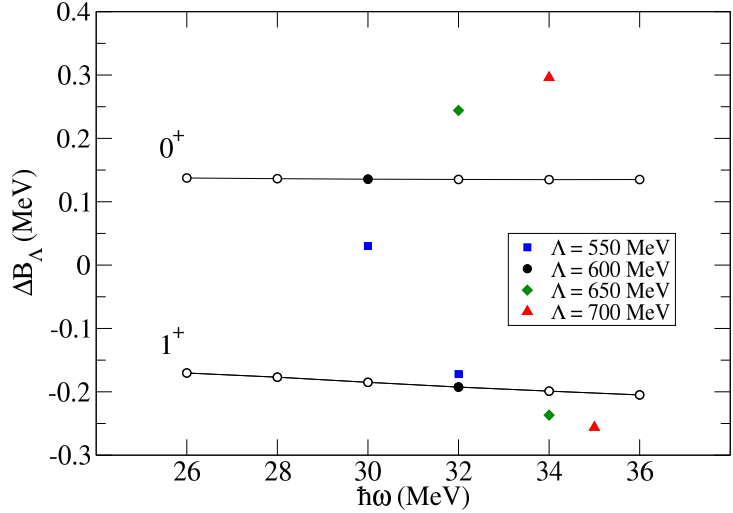} 
\caption{Dependence of the separation-energy differences $\Delta B_{\Lambda}$ 
between $_{\Lambda}^4{\rm He}$ and $_{\Lambda}^4{\rm H}$, for $0^+_{\rm g.s.}$ 
(upper curve) and for $1^+_{\rm exc}$ (lower curve) on the HO $\hbar\omega$ 
in {\it ab initio} NCSM calculations using LO chiral EFT coupled-channel $YN$ 
potentials with cutoff momentum $\Lambda$=600~MeV \cite{polinder06} plus 
$V_{\rm CSB}$ derived from these SI potentials using Eq.~(\ref{eq:OME}). 
Results for other values of $\Lambda$ are shown at the respective absolute 
variational energy minima.} 
\label{fig:csb} 
\end{center} 
\end{figure}  

In Fig.~\ref{fig:csb} we show the $\omega$ dependence of separation-energy 
differences $\Delta B^{J}_{\Lambda}$ between $_{\Lambda}^4{\rm He}$ and 
$_{\Lambda}^4{\rm H}$ levels of a given spin $J$, for $0^+_{\rm g.s.}$ 
and $1^{+}_{\rm exc.}$, using $N_{\rm max}\to\infty$ extrapolated values 
for the four possible binding energies which are calculated for a cutoff 
$\Lambda$=600~MeV and including $V_{\rm CSB}$ from Eq.~(\ref{eq:OME}). 
Extrapolation uncertainties for $\Delta B^{J}_{\Lambda}$ are between 10 to 
20~keV. The variation of $\Delta B^{J=0}_{\Lambda}$ in the $\hbar\omega$ 
range spanned in the figure amounts to a few keV, whereas that of $\Delta 
B^{J=1}_{\Lambda}$ is larger, amounting to $\sim$30~keV. It is worth noting 
that the difference $\Delta B^{J=0}_{\Lambda}-\Delta B^{J=1}_{\Lambda}$ 
between the upper and lower curves assumes at $\Lambda$=600~MeV the value 
0.33$\pm$0.04~MeV, in perfect agreement with the difference $E_{\gamma}(_{
\Lambda}^{4}{\rm He})-E_{\gamma}(_{\Lambda}^{4}{\rm H})=0.32\pm 0.02$~MeV 
between the two $\gamma$ ray energies shown in Fig.~\ref{fig:A=4}. The figure 
also shows, again, a strong cutoff dependence of $\Delta B^{J=0}_{\Lambda}$ 
together with a moderate cutoff dependence of $\Delta B^{J=1}_{\Lambda}$.

\section{Discussion} 
\label{sec:disc} 

Dalitz and von Hippel \cite{DvH64} who suggested the CSB $\Lambda - \Sigma^0$ 
mass-mixing mechanism, realized its great merit of generating a $\Lambda N$ 
OPE long-range CSB potential, $V_{\rm CSB}^{\rm OPE}$, otherwise forbidden by 
the strong interactions. Disregarding tensor components of OPE, DvH estimated 
$\Delta B^{J=0}_{\Lambda}({\rm OPE})=165$~keV for the $A$=4 hypernuclear g.s. 
Updating some of the relevant coupling constants, their $0^+_{\rm g.s.}$ 
wave function yields $\Delta B^{J=0}_{\Lambda}({\rm OPE})\approx 95$~keV. 
An exploratory four-body $\Lambda NNN$ Monte Carlo calculation limited to 
relative $S$ states by Coon et al. \cite{coon99} yielded a smaller value 
of $\Delta B^{J=0}_{\Lambda}({\rm OPE})\sim 45$~keV, augmented though by 
a larger contribution from the very short-ranged $\rho$ meson exchange, 
$\Delta B^{J=0}_{\Lambda}({\rm ORE})\sim 75$~keV. The first $YNNN$ 
coupled-channel four-body calculation of the $A$=4 hypernuclei 
\cite{nogga01,nogga02}, using the coupled-channel $YN$ interaction 
models NSC97 \cite{NSC97}, incorporated OPE and ORE CSB contributions, 
including those from tensor-interaction components, as well perhaps as other 
contributions. Surprisingly, small values of $\Delta B^{J=0}_{\Lambda}$ 
were found, about 75~keV \cite{nogga02} and 100~keV \cite{haidenbauer07} 
in versions e and f, respectively, of NSC97. 

All of the calculations mentioned above agree in sign, $\Delta B^{J=0}_{
\Lambda}({\rm OPE})>0$, with the experimentally derived value of $\Delta B^{
J=0}_{\Lambda}$. However, in the present calculations, a negative OPE 
contribution is indicated by Eq.~(\ref{eq:MO0}). To understand this apparent 
disagreement we list in Tables~\ref{tab:OPEgs550} and \ref{tab:OPEgs600} 
partial $V_{\rm CSB}^{\rm OPE}$ contributions to $\Delta B^{J=0}_{\Lambda}$, 
computed by adding $V_{\rm CSB}^{\rm OPE}$ directly to the LO $\chi$EFT 
coupled-channel $YN$ potential $V_{\rm SI}$, without activating 
Eq.~(\ref{eq:OME}) which relates $V_{\rm CSB}$ to $V_{\rm SI}$. The SI 
potentials $V_{\rm SI}$ were regularized by using a cutoff momentum 
$\Lambda_{\rm SI}=550$~MeV in Table~\ref{tab:OPEgs550} and $\Lambda_{\rm SI}
=600$~MeV in Table~\ref{tab:OPEgs600}, whereas the momentum-space 
$V_{\rm CSB}^{\rm OPE}$ was regularized using a sequence of 
$\Lambda_{\rm CSB}$ values, $\Lambda_{\rm CSB}=600,700$~MeV, 
in each one of the tables. Similarly, results for 
$\Delta B^{J=1}_{\Lambda}({\rm OPE})$ are listed in Tables~\ref{tab:OPEexc550} 
and \ref{tab:OPEexc600} below. We also checked the limit 
$\Lambda_{\rm CSB}\to\infty$, in which $V_{\rm CSB}^{\rm OPE}$ 
is not regularized. 

\begin{table}[hbt]
\caption{OPE partial (central and tensor) contributions (in keV) to $\Delta
B^{J=0}_{\Lambda}$ in NCSM $A$=4 binding energy calculations, using the
Bonn-J\"{u}lich LO $\chi$EFT SI $YN$ potentials \cite{polinder06} with cutoff 
$\Lambda_{\rm SI}=550$~MeV. The CSB OPE potential is regularized using cutoff 
values $\Lambda_{\rm CSB}$ (in MeV). The limiting case $\Lambda_{\rm CSB}\to
\infty$ corresponds to unregularized CSB OPE potential. For the meaning of 
the DvH entries, see text.}
\begin{tabular}{lccccc}
\hline
$\Lambda_{\rm CSB}$ & central & tensor & LO $\chi$EFT & central DvH &
updated DvH \\
\hline
600 & $-$298 & $+$37  & $-$224 & $+$109 & $+$146 \\
700 & $-$311 & $+$55 & $-$218 & $+$108 & $+$163 \\
$\Lambda_{\rm CSB}\to\infty$ & $-$329 & $+$88 & $-$203 & $+$110 & $+$198 \\
\hline
\end{tabular}
\label{tab:OPEgs550}
\end{table}

\begin{table}[hbt]
\caption{Same as Table~\ref{tab:OPEgs550}, but for $\Lambda_{\rm SI}=600$~MeV 
instead of 550~MeV.} 
\begin{tabular}{lccccc}
\hline 
$\Lambda_{\rm CSB}$ & central & tensor & LO $\chi$EFT & central DvH & 
updated DvH \\
\hline 
600 & $-$264 & $+$81  & $-$167 & $+$102 & $+$183 \\ 
700 & $-$277 & $+$107 & $-$155 & $+$104 & $+$211 \\ 
$\Lambda_{\rm CSB}\to\infty$ & $-$297 & $+$158 & $-$124 & $+$106 & $+$264 \\ 
\hline 
\end{tabular} 
\label{tab:OPEgs600} 
\end{table} 

The OPE potential has two components with contributions listed in the second 
and third columns: (i) a spin-dependent central component and (ii) a tensor 
component. These two partial contributions add up approximately, taking 
into account the `background CSB' contributions of Table~\ref{tab:CS} in 
Sect.~\ref{subsec:noCSB}, to the summed OPE contribution in the LO $\chi$EFT 
interaction model given in the fourth column. A spin dependence ${\vec 
\sigma}_{\Lambda}\cdot {\vec \sigma}_N$ is responsible for the approximate 
ratio $-$3:1 of the $J=0$ to $J=1$ central contributions. However, these 
contributions are of opposite sign to those expected naively from OPE. The 
resolution of the puzzle is that the central component of this OPE potential, 
like all PS ${\bf 8}_{\rm f}$ exchange potentials in the Bonn-J\"{u}lich 
model, consists of two opposite-sign terms which in coordinate space are the 
familiar Yukawa exponential potential of range $m_{\pi}^{-1}$ and a Dirac 
$\delta ({\vec r})$ zero-range potential. Because both have the same volume 
integral, the contribution of the $\delta({\vec r})$ piece is larger in 
magnitude than the Yukawa contribution, even when smeared by the regularizing 
form factors, and this is how the sign of the central contribution (second 
column) in the tables is opposite to what DvH anticipated. Removing the 
smeared $\delta({\vec r})$ term from $V_{\rm CSB}^{\rm OPE}$, one reverses 
the sign of the central contribution, with the modified central contribution 
listed in the fifth column under `central DvH'. As deduced from 
Tables~\ref{tab:OPEgs550}, \ref{tab:OPEgs600}, \ref{tab:OPEexc550} 
and \ref{tab:OPEexc600}, this contribution is insensitive to any of the two 
cutoffs, $\Lambda_{\rm SI}$ and $\Lambda_{\rm CSB}$, within the range of 
values varied, and the corresponding $\approx$105~keV contribution for $J=0$ 
is consistent with the rough update mentioned above of the DvH central-OPE 
estimate. In contrast, the tensor contribution, particularly for $J=0$, 
is more sensitive to each one of the cutoffs, with a spread of values from 
the finite $\Lambda_{\rm CSB}$ entries in Tables~\ref{tab:OPEgs550} and 
\ref{tab:OPEgs600} given by $\sim$70$\pm$35~keV in the $0^+_{\rm g.s.}$. 
Altogether the `updated DvH' total OPE CSB contribution to $\Delta B^{J=0}_{
\Lambda}$ inferred from the finite $\Lambda_{\rm CSB}$ rows is quite large, 
$\sim$175$\pm$40~keV, with a much smaller-size and negative total OPE CSB 
contribution, $\approx -$48$\pm$10~keV, to $\Delta B^{J=1}_{\Lambda}$. 
This would fit remarkably well the observed CSB splittings.{\footnote{If 
$\Lambda_{\rm SI}$ values of 600 and 650~MeV that are the closest ones to 
reproducing the observed $E_{\rm x}$ values, see Fig.~\ref{fig:Ex}, are used 
instead, the `updated DvH' total OPE CSB contribution to $\Delta B^{J=0}_{
\Lambda}$ increases to $\sim$235$\pm$25~keV and that to $\Delta B^{J=1}_{
\Lambda}$ slightly changes to $\approx -$35$\pm$9~keV.}}  
 
\begin{table}[hbt]
\caption{Same as Table~\ref{tab:OPEgs550}, but for $\Delta B^{J=1}_{\Lambda}$.}
\begin{tabular}{lccccc}
\hline
$\Lambda_{\rm CSB}$ & central & tensor & LO $\chi$EFT & central DvH &
updated DvH \\
\hline
600 & $+$60 & $-$17 & $+$95 & $-$40 & $-$57 \\
700 & $+$68 & $-$12 & $+$108 & $-$40 & $-$52 \\
$\Lambda_{\rm CSB}\to\infty$ & $+$79 & $-$2 & $+$129 & $-$42 & $-$44 \\
\hline
\end{tabular}
\label{tab:OPEexc550}
\end{table}

\begin{table}[hbt]
\caption{Same as Table~\ref{tab:OPEgs600}, but for $\Delta B^{J=1}_{\Lambda}$.} 
\begin{tabular}{lccccc}
\hline 
$\Lambda_{\rm CSB}$ & central & tensor & LO $\chi$EFT & central DvH & 
updated DvH \\
\hline 
600 & $+$73 & $-$4 & $+$109 & $-$39 & $-$43 \\ 
700 & $+$82 & $+$2 & $+$127 & $-$40 & $-$38 \\ 
$\Lambda_{\rm CSB}\to\infty$ & $+$96 & $+$15 & $+$151 & $-$42 & $-$27 \\
\hline 
\end{tabular} 
\label{tab:OPEexc600} 
\end{table} 

For the finite values of the cutoff $\Lambda_{\rm CSB}$ listed in Tables 
\ref{tab:OPEgs550}, \ref{tab:OPEgs600}, \ref{tab:OPEexc550} and 
\ref{tab:OPEexc600}, the dependence of the CSB OPE contributions on 
$\Lambda_{\rm CSB}$ for a given $\Lambda_{\rm SI}$ is weak to moderate, 
and the limiting case of $\Lambda_{\rm CSB}\to\infty$ poses no convergence 
problem. However, once $\Lambda_{\rm CSB}$ is increased beyond roughly 
700~MeV, the ORE contribution may no longer be ignored, with a $\delta({
\vec r})$-subtracted central contribution that augments the OPE $\delta({
\vec r})$-subtracted central contribution and a tensor contribution that 
reduces considerably the OPE tensor contribution. We conjecture that 
the failure of the NSC97 $YN$ models to reproduce the large size of the 
observed g.s.~CSB splitting $\Delta B^{J=0}_{\Lambda}$ arises from a strong 
cancellation between the OPE and ORE tensor CSB contributions which in these 
models overshadow the central CSB contributions. 

Finally, the dependence of the total, `updated DvH' OPE CSB contribution on 
the strong-interaction cutoff $\Lambda_{\rm SI}$, for a given $\Lambda_{
\rm CSB}$, is considerably weaker for the $0^+_{\rm g.s.}$ than that given 
in Table~\ref{tab:CSB} using Eq.~(\ref{eq:OME}) to derive $V_{\rm CSB}$.

\section{Summary and outlook} 
\label{sec:sum} 

In this work we discussed the extension of the NCSM from few-body nuclear 
to few-body hypernuclear applications and provided details of our recent 
Letter publication on {\it ab initio} calculations of CSB in the $A$=4 
mirror hypernuclei \cite{gg16}. These calculations are the first microscopic 
calculations to generate a large positive value of $\Delta B^{J=0}_{\Lambda}$ 
commensurate with experiment, although with a considerable momentum-cutoff 
dependence within the Bonn-J\"{u}lich LO $\chi$EFT coupled-channel $YN$ 
potential model \cite{polinder06}. The calculational extrapolation 
uncertainties involved in the evaluation of $\Delta B^{J=0}_{\Lambda}$ were 
estimated to be in the range of 10 to 20 keV at most. In the Bonn-J\"{u}lich 
$\chi$EFT approach, the relatively large value derived for $\Delta B^{J=0}_{
\Lambda}$ arises from the $^1S_0$ CT of the SI $\Lambda N\leftrightarrow\Sigma 
N$ coupling potential, appearing to have no relationship with the large OPE 
CSB contribution anticipated by DvH \cite{DvH64}. This is a direct consequence 
of using the relationship given by Eq.~(\ref{eq:OME}) between SI and CSB. 
By removing the short-range $\delta ({\vec r})$ term from the OPE $\Lambda N$ 
CSB potential, and using a DWBA-like evaluation of this CSB potential, we were 
able to recover the DvH original estimate of the central OPE CSB contribution, 
updated to present-day coupling constants. Furthermore, choosing a cutoff 
$\Lambda_{\rm SI}$=600~MeV, which is closer to reproducing $E_{\rm x}(0^+_{
\rm g.s.}\to 1^{+}_{\rm exc})_{\rm exp}$ than the lower cutoff considered 
here, as large values as $\sim$200~keV for $\Delta B^{J=0}_{\Lambda}$, and 
small and negative values $\approx -$40~keV for $\Delta B^{J=1}_{\Lambda}$, 
emerge for the combined central plus tensor OPE CSB contribution, in striking 
agreement with experiment. Similar estimates were obtained by one of the 
authors \cite{gal15} using a $\Lambda N \leftrightarrow \Sigma N$ effective 
$V_{\rm SI}$ to which Eq.~(\ref{eq:OME}) was applied to generate the 
corresponding $V_{\rm CSB}$. 

Future applications of the NCSM to $p$-shell hypernuclei are desirable, 
in view of the few CSB mirror-level splittings known in this mass range 
\cite{gal15}. The lesson of this latter work is that genuine CSB splittings 
become smaller as one goes to heavier hypernuclei. In this respect, given 
the particularly large observed value of $\Delta B^{J=0}_{\Lambda}$ in the 
$A$=4 mirror hypernuclei considered in the present work, these hypernuclei 
provide a unique test ground for CSB models beyond nuclear physics.

\section*{Appendix A: Jacobi-coordinate NCSM hypernuclear applications} 
\label{sec:appendix} 
\renewcommand{\theequation}{A.\arabic{equation}}
\setcounter{equation}{0}
\renewcommand{\thefigure}{A.\arabic{figure}} 
\setcounter{figure}{0} 

The starting point of the \textit{ab initio} NCSM calculations is the 
Hamiltonian for a system of nonrelativistic nucleons and hyperons 
interacting by realistic two-body $NN$ and $YN$, and also three-nucleon 
interactions:  
\begin{equation}
\label{eq:h}
H=\sum_{i=1}^A \frac{\vec{p}^{\,2}_{i}}{2m_i}+\sum_{i<j=1}^A
V(\vec{r}_i,\vec{r}_j) + \sum_{i<j<k=1}^{A-1}
V(\vec{r}_i,\vec{r}_j,\vec{r}_k).
\end{equation}
In the present work, considering the $A$=4 mirror hypernuclei, the momenta 
$\vec{p}_i$, masses $m_i$ and coordinates $\vec{r}_i$ for $i=1,2,3$ correspond 
to nucleons and those for $i=4$ to hyperons. The Hamiltonian form (\ref{eq:h}) 
is then rewritten in terms of {\it relative} Jacobi coordinates, momenta 
and their associated masses. There are several different sets of Jacobi 
coordinates, The first of which is defined by 
\begin{align}
\label{eq:cset1}
\begin{split}
\vec{\xi}_0 &= \sqrt{\frac{1}{M}}
\sum_{i=1}^4 m_i \vec{r}_i, \\
\vec{\xi}_1 &= \sqrt{\frac{m_1m_2}{m_1+m_2}} (\vec{r}_1-\vec{r}_2),\\
\vec{\xi}_2 &=
\sqrt{\frac{(m_1+m_2)m_3}{m_1+m_2+m_3}}\left(\frac{m_1\vec{r}_1+m_2\vec{r}_2}
{m_1+m_2}-\vec{r}_3\right),\\
\vec{\xi}_3 &=
\sqrt{\frac{(m_1+m_2+m_3)m_4}{M}}
\left(\frac{m_1\vec{r}_1+m_2\vec{r}_2+m_3\vec{r}_3}{m_1+m_2+m_3}-\vec{r}_4
\right),
\end{split}
\end{align}
where $M=\sum_{i=1}^4m_i$. This particular set is a natural one for 
implementing antisymmetrization with respect to nucleons, and is subsequently 
used for diagonalization of the Hamiltonian. Here, $\vec{\xi}_0$ is 
proportional to the center of mass coordinate of the $A$-baryon system and 
$\vec{\xi}_i$ ($i>0$) is proportional to the relative coordinate of the 
$i+1$ baryon with respect to the center of mass of $\leq i$ baryons. 
The kinetic energy term in Eq.~(\ref{eq:h}) is then rewritten in terms of  
Jacobi cooredinates (\ref{eq:cset1}): 
\begin{equation}
\label{eq:ht}
\sum_{i=1}^4 \frac{\vec{p}^{\,2}_{i}}{2m_i}\equiv
-\sum_{i=1}^4
\frac{1}{2m_i}\vec{\nabla}^{\,2}_{\vec{r}_i}=
-\frac{1}{2}\vec{\nabla}^{\,2}_{\vec{\xi}_0}
-\sum_{i=1}^3\frac{1}{2}\vec{\nabla}^{\,2}_{\vec{\xi}_i}.
\end{equation}
Since the various interactions $V$ in (\ref{eq:h}) do not depend on 
$\vec{\xi}_0$, the center of mass kinetic energy can be omitted from 
(\ref{eq:ht}), and one can use an HO basis depending on coordinates 
$\vec{\xi}_1$, $\vec{\xi}_2$ and $\vec{\xi}_3$, e.g.\ 
\begin{equation}
\label{eq:b1}
\vert ((nlsjt)n_3l_3j_3)J_NT_N,n_Yl_Yj_Yt_Y)JT\rangle.
\end{equation}
Here $n$, $l$ are HO quantum numbers corresponding to coordinate $\vec{\xi}_1$ 
describing the relative motion of the first two nucleons; $n_3$, $l_3$ 
corresponding to $\vec{\xi}_2$ describe the relative motion of the third 
nucleon with respect to the nucleon pair; and $n_Y$, $l_Y$ associated with 
$\vec{\xi}_3$ describe the relative motion of the hyperon with respect to the 
three-nucleon cluster. The spin quantum numbers referring to single-particle 
states are omitted, $s=0,1$ is the spin of the two-nucleon pair, and the $j$ 
quantum numbers denote respective angular momenta. We work in the isospin 
basis, $t=0,1$ is the isospin of the nucleon pair, and the nucleon 
single-particle isospin is also suppressed in (\ref{eq:b1}). The hyperon 
isospin quantum number $t_Y=0,1$ holds for $\Lambda$ and $\Sigma$ hyperons, 
respectively, thereby allowing for explicit admixtures of $\Sigma$ hyperons 
into $\Lambda$ hypernuclear states, induced by the $t_{YN}=\frac{1}{2}$ 
$\Lambda N\leftrightarrow\Sigma N$ coupling potential. The basis (\ref{eq:b1}) 
is truncated in NCSM calculations by requiring that the total number of HO 
quanta does not exceed a chosen value $N_\text{max}$, 
\begin{equation}
2n+l+2n_3+l_3+2n_Y+l_Y \le N_\text{max}
\end{equation}
thereby defining the size of the model space. Moreover, all HO wave functions 
in (\ref{eq:b1}) depend on a single HO frequency $\omega$ which is a free 
parameter in NCSM calculations.

The basis (\ref{eq:b1}) is antisymmetric with respect to exchanging nucleons 
$1$ and $2$ upon requiring $(-1)^{l+s+t}=-1$ for the two-nucleon system. 
It is, however, not antisymmetric with respect to nucleon exchanges $1 
\leftrightarrow 3$ and $2 \leftrightarrow 3$. The procedure of fully 
antisymmetrizing the three-nucleon cluster in the basis (\ref{eq:b1}), 
recalling that it is disconnected from the hyperon quantum numbers, is 
described in detail e.g.\ in Ref.~\cite{navratil00}. The resulting fully 
antisymmetric three-nucleon cluster basis elements can be expanded as linear 
combinations of the original basis (\ref{eq:b1}). Incidentally, the set of 
coordinates (\ref{eq:cset1}) is also suitable for evaluating three-nucleon 
interaction matrix elements which are naturally expressed as functions of 
the Jacobi coordinates $\vec{\xi}_1$ and $\vec{\xi}_2$ \cite{navratil07}.

The basis (\ref{eq:b1}) is, however, inappropriate for evaluating two-body 
interaction terms. Another set of Jacobi coordinates suitable for basis 
expansion when $NN$ and $YN$ interaction matrix elements are calculated is 
obtained by keeping to $\vec{\xi}_0$, $\vec{\xi}_1$ and introducing two new 
variables, 
\begin{align}
\label{eq:cset2}
\begin{split}
\vec{\eta}_{2} &= \sqrt{\frac{(m_1+m_2)(m_3+m_4)}{M}}
\left(\frac{m_1\vec{r}_1+m_2\vec{r}_{2}}{m_1+m_2}
-\frac{m_3\vec{r}_3+m_4\vec{r}_{4}}{m_3+m_4}\right),\\
\vec{\eta}_{3} &= \sqrt{\frac{m_3m_4}{m_3+m_4}}
\left(\vec{r}_3-\vec{r}_4\right).
\end{split}
\end{align}
A basis depending on coordinates $\vec{\xi}_1$, $\vec{\eta}_1$, 
$\vec{\eta}_2$, with two-body subclusters, may be defined e.g.\ as 
\begin{equation}
\label{eq:b2}
\vert ((nlsjt),(n_{YN}l_{YN}s_{YN}j_{YN}t_{YN},\mathcal{N}\mathcal{L})
\mathcal{J})JT\rangle, 
\end{equation} 
where, similarly to (\ref{eq:b1}), the HO state $\vert nlsjt \rangle$ 
associated with the coordinate $\vec{\xi}_1$ describes the nucleon pair and 
the HO state $\vert n_{YN}l_{YN}s_{YN}j_{YN}t_{YN}\rangle$ associated with 
$\vec{\eta}_3$ corresponds to the relative-coordinate hyperon--nucleon 
channel, with $s_{YN}=0,1$, $j_{YN}$ and $t_{YN}=\frac{1}{2},\frac{3}{2}$ 
standing for the spin, total angular momentum and isospin of the $YN$ pair, 
respectively. The HO state $\vert\mathcal{N}\mathcal{L}\rangle$ associated 
with the coordinate $\vec{\eta}_2$ describes the relative motion of the $NN$ 
and $YN$ clusters. Properties of HO wave functions and Jacobi coordinates 
allow basis elements defined in (\ref{eq:b1}) to be expanded in basis 
(\ref{eq:b2}) as follows: 
\begin{align}
\label{eq:rcpl}
\begin{split}
&\vert ((nlsjt)n_3l_3j_3)J_NT_N,n_Yl_Yj_Yt_Y)JT\rangle \\
&=\sum 
\hat{T}_{N} \hat{t}_{YN} (-1)^{t+\tfrac{1}{2}+t_Y+T}
\begin{Bmatrix}
t & \tfrac{1}{2} & T_N \\
t_Y & T & t_{YN}
\end{Bmatrix}
\\
&\times \hat{j}_Y \hat{J}_N \hat{L}^2\hat{j}_3 \hat{s}_{YN}
\hat{\mathcal{J}}\hat{j}_{YN}
(-1)^{j+j_3+J_N+J+\mathcal{L}+j_{YN}+l_3+l_Y+s_{YN}}
\\
&\times 
\begin{Bmatrix}
l_{3} & \tfrac{1}{2} & j_3 \\
l_Y & \tfrac{1}{2} & j_Y \\
L & s_{YN} & \mathcal{J}
\end{Bmatrix}
\begin{Bmatrix}
j & j_3 & J_N \\
j_Y & J & \mathcal{J}
\end{Bmatrix}
\begin{Bmatrix}
\mathcal{L} & l_{YN} & L \\
s_{YN} & \mathcal{J} & j_{YN}
\end{Bmatrix}
\\
&\times \langle n_{YN}l_{YN}\mathcal{NL}L 
\vert n_Yl_Yn_3l_3L \rangle_{\tfrac{3m+m_Y}{2m_Y}} \\
&\vert ((nlsjt),
(n_{YN}l_{YN}s_{YN}j_{YN}t_{YN},\mathcal{N}\mathcal{L})\mathcal{J})JT\rangle,
\end{split}
\end{align}
where the orthogonal transformation between the Jacobi coordinates 
$\vec{\xi}_{2},\vec{\xi}_{3}$ and $\vec{\eta}_{2},\vec{\eta}_{3}$ was employed, 
and $\langle n_{YN}l_{YN}\mathcal{NL}L\vert n_Yl_Yn_3l_3L\rangle_{
\tfrac{3m+m_Y}{2m_Y}}$ is the general HO bracket for two particles, 
defined e.g\ in Ref.~\cite{trlifaj73}. Here, $m$ and $m_Y$ are the 
nucleon and hyperon ($Y=\Lambda$, $\Sigma$) masses defined as 
\begin{align}
m &= \frac{m_n+m_p}{2}+\frac{m_n-m_p}{A}M_T, \\
m_\Sigma &= \frac{m_{\Sigma^-}+m_{\Sigma^0}+m_{\Sigma^+}}{3},
\end{align}
with $m_n$, $m_p$, $m_{\Sigma^-}$, $m_{\Sigma^0}$, and $m_{\Sigma^+}$ denoting 
the masses of the neutron, proton, $\Sigma^-$, $\Sigma^0$, and $\Sigma^+$ 
hyperons, respectively, and $M_T$ is the projection of the total isospin T, 
$M_T=\mp \frac{1}{2}$ for (\lamb{4}{H},\,\lamb{4}{He}) respectively. 
The transformation (\ref{eq:rcpl}) conserves the total $J$ and $T$ and also, 
quite importantly, the total number of HO quanta, 
\begin{equation}
2n+l+2n_3+l_3+2n_Y+l_Y = 2n+l+2n_{YN}+l_{YN}+2\mathcal{N}+\mathcal{L}.
\end{equation}
Using the expansion (\ref{eq:rcpl}), it is straightforward to evaluate matrix 
elements of two-body interactions in the basis (\ref{eq:b1}), 
\begin{align}
\label{eq:v2nn}
\langle \sum_{i<j =1}^{3} V_{ij} \rangle &=
3\langle V_{NN}(\sqrt{\tfrac{2}{m}}\vec{\xi}_1)\rangle,\\
\label{eq:v2ny}
\langle \sum_{i=1}^{3} V_{i4} \rangle &=
3\langle V_{YN}(\sqrt{\tfrac{m+m_Y}{m\,m_Y}}\vec{\eta}_{3})
\rangle,
\end{align}
where the matrix elements on the right hand sides are diagonal in all 
quantum numbers of the states (\ref{eq:b2}) except for $n$, $l$ and 
$n_{YN}$, $l_{YN}$, respectively, for isospin conserving interactions. 
Equally straightforward is the evaluation of two-body interactions defined in 
momentum space, since transformations analogous to those in (\ref{eq:cset1}) 
and (\ref{eq:cset2}) can be introduced for momenta $\vec{p}_i$ by substituting 
$\vec{r}_i \rightarrow\frac{\vec{p}_i}{m_i}$. Both local and non-local 
interactions can be accommodated within the NCSM methodology. 

Realistic $NN$ and $YN$ interactions are, however, usually defined in the 
particle basis, not in the isospin basis. To evaluate the corresponding matrix 
elements of $V_{NN}$ between good-isospin basis states (\ref{eq:b2}) we use 
the following prescription
\begin{equation}
\label{eq:vnnavg}
\begin{split}
&\langle (t',t_{YN}')TM_T\vert V_{NN} \vert (t,t_{YN})TM_T \rangle
=\delta_{t't} \delta_{t_{YN}'t_{YN}} \\
&\times\sum \langle t\,m\,t_{YN}\,m_{YN} \vert T\, M_T\rangle^2\\
&\times \langle \tfrac{1}{2}\, m_1'\,\tfrac{1}{2}\,m_2' \vert t\, m \rangle
\langle \tfrac{1}{2}\, m_1\,\tfrac{1}{2}\,m_2 \vert t\, m \rangle \\
&\times \langle \tfrac{1}{2}\, m_1',\tfrac{1}{2}\,m_2' \vert V_{NN} \vert
\tfrac{1}{2}\, m_1,\tfrac{1}{2}\,m_2 \rangle \\
&\equiv V_{NN}(t;t_{YN},T,M_T).
\end{split}
\end{equation}
Here, only the isospin quantum numbers of states (\ref{eq:b2}) are displayed. 
The basis elements are decomposed via Clebsch--Gordan coefficients and the 
potential matrix elemets are evaluated between two-nucleon states $\vert 
\tfrac{1}{2}\,m_1,\tfrac{1}{2}\,m_2 \rangle$ with single-nucleon isospin 
projections $m_1=\pm\frac{1}{2}$ and $m_2=\pm\frac{1}{2}$. In this procedure 
the isospin breaking transitions $t=0 \leftrightarrow 1$ are suppressed, but 
the resulting isospin-basis defined $NN$ interaction depends parametrically 
on the isospin of the $YN$ cluster, as well as on the total isospin and its 
projection. Similarly, a particle-basis defined $YN$ interaction $V_{YN}$ 
is evaluated as 
\begin{equation}
\label{eq:vnyavg}
\begin{split}
&\langle (t',t_{YN}')TM_T\vert V_{YN} \vert (t,t_{YN})TM_T \rangle
=\delta_{t't} \delta_{t_{YN}'t_{YN}} \\
&\times\sum \langle t\,m\,t_{YN}\,m_{YN} \vert T\, M_T\rangle^2\\
&\times
\langle 1\,m_1'\,\tfrac{1}{2}\, m_2' \vert t_{YN}\, m_{YN} \rangle
\langle 1\,m_1 \,\tfrac{1}{2}\, m_2  \vert t_{YN}\, m_{YN} \rangle \\
&\times \langle 1\,m_1', \tfrac{1}{2}, m_2' \vert V_{YN} \vert
1\,m_1, \tfrac{1}{2}\, m_2 \rangle \\
&\equiv V_{YN}(t_{YN};t,T,M_T),
\end{split}
\end{equation} 
where the potential matrix elements are evaluated between hyperon--nucleon 
states $\vert 1\, m_1, \tfrac{1}{2}\, m_2\rangle$ with $m_1=-1,0,1$ and 
$m_2=\pm\frac{1}{2}$ the isospin projections of hyperon $Y$ and nucleon $N$, 
respectively. Again, the isospin-breaking transitions $t_{YN}=\frac{1}{2}
\leftrightarrow\frac{3}{2}$ are suppressed. This procedure gives 
excellent agreement with particle-basis calculations as demonstrated in 
Ref.~\cite{wirth14}. For the $A$=3,4 hypernuclear systems, the difference 
between calculated total energies in particle basis and isospin basis using 
relations (\ref{eq:vnnavg}) and (\ref{eq:vnyavg}) was found to be only few 
keV.

\section*{Acknowledgments} 

We are grateful to Petr Navr\'{a}til for helpful advice on extensions of 
nuclear-physics NCSM codes, to Johann Haidenbauer, and Andreas Nogga for 
providing us with the input LO EFT $YN$ potentials used in the present work, 
and to Nir Barnea, Ji\v{r}\'{i} Mare\v{s} and Ulf Mei{\ss}ner for useful 
discussions on issues related to this work. The research of D.G. was supported 
by the Granting Agency of the Czech Republic (GACR), Grant No. P203/15/04301S.

\end{document}